%Paper: astro-ph/9505017
%From: Zuhui Fan <fanz@fermi.phys.washington.edu>
%Date: Thu, 4 May 1995 09:36:52 -0700 (PDT)

%file wbasict.tex (170 lines)
\magnification=1000
\hoffset=1.2cm \voffset=0.2cm
\vbadness=10000

\font\SS=cmss10 scaled 1200

\font\ss=cmssq8 scaled 1200
\font\bf=cmbx10 scaled 1200
\font\bb=cmbx10 scaled 1728

\font\rm=cmr10 scaled 1440

\font\it=cmti10 scaled 1200

\font\sc=cmcsc10 scaled 1200
\font\tenrm=cmr10 scaled 1200
\font\sevenrm=cmr7 scaled 1000
\font\fiverm=cmr5 scaled 1000
\font\teni=cmmi10 scaled 1200
\font\seveni=cmmi7 scaled 1000
\font\fivei=cmmi5 scaled 1000
\font\sevensy=cmsy7 scaled 1000
\font\tensy=cmsy10 scaled 1200
\font\fivesy=cmsy5 scaled 1000

\font\tenbf=cmbx10 scaled 1200
\font\sevenbf=cmbx7 scaled 1200
\font\fivebf=cmbx5 scaled 1200
\font\tensl=cmsl10 scaled 1200
\font\tentt=cmtt10 scaled 1200
\font\tenit=cmti10 scaled 1200
\catcode`\@=11
\textfont0=\tenrm \scriptfont0=\sevenrm \scriptscriptfont0=\fiverm
\def\rm{\fam\z@\tenrm}
\textfont1=\teni \scriptfont1=\seveni \scriptscriptfont1=\fivei
\def\mit{\fam\@ne} \def\oldstyle{\fam\@ne\teni}
\textfont2=\tensy \scriptfont2=\sevensy \scriptscriptfont2=\fivesy
\def\cal{\fam\tw@}
\textfont3=\tenex \scriptfont3=\tenex \scriptscriptfont3=\tenex
\newfam\itfam \def\it{\fam\itfam\tenit} % \it is family 4
\textfont\itfam=\tenit
\newfam\slfam  % \sl is family 5
\textfont\slfam=\tensl
\newfam\bffam \def\bf{\fam\bffam\tenbf} % \bf is family 6
\textfont\bffam=\tenbf \scriptfont\bffam=\sevenbf
\scriptscriptfont\bffam=\fivebf
\newfam\ttfam  % \tt is family 7
\textfont\ttfam=\tentt
\catcode`\@=12
\rm

\hfuzz=10pt \overfullrule=0pt
\vsize=8.9in
\hsize=14.9cm
\baselineskip=12pt
\def\doublespace{\baselineskip=30pt}
\def\singlespace{\baselineskip=12pt}
\parindent 20pt \parskip 6pt

\def\ctrline#1{\centerline{#1}}
\def\linebreak{\hfil\break}
\def\blankline{\par\vskip \baselineskip}
\def\twocol#1{\halign{##\quad\hfil &##\hfil\cr #1}}

 \mathcode`*="002A

\def\neq{\not=}
\def\<={\leq}
\def\>={\geq}
\def\lsls{\ll}

\def\^{\char'017{}}
\def\%{\char'045{}}
\def\_{\vrule height 0.8pt depth 0pt width 1em}

\let\Mdollar=\$
\def\${\ifmmode\Mdollar\else$\Mdollar$\fi}

\def\spose\rlap
%def\simless{\mathbin{\lower 1pt\hbox
%  {$\spose{\raise 5pt\hbox{$\char'074$}}\char'430$}}}
%def\simgreat{\mathbin{\lower 1pt\hbox
%  {$\spose{\raise 5pt\hbox{$\char'076$}}\char'430$}}}
\def\caret#1{\widehat #1}
\def\tilde#1{\widetilde #1}

\def\cfalh{\par\vfil\eject \vskip -12pt \moveleft 0.5in\vbox{
     \twocol{{\bb Center for Astrophysics}\hbox to 1.5in{} &\cr
     {\ss 60 Garden Street} & {\ss Harvard College Observatory} \cr
     {\ss Cambridge, Massachusetts 02138} & {\ss Smithsonian
          Astrophysical Observatory}\cr}}\par\blankline}
\def\physicslh{\par\vfil\eject \vskip -12pt \vbox{
     \ctrline{\SS HARVARD UNIVERSITY}
     \vbox to 5 pt {}
     \hbox to \hsize{{\ss Department of
Physics}\hfil {\ss Lyman Laboratory
          of Physics}}
     \hbox to \hsize{\hfil {\ss Cambridge, Massachusetts
02138}}}\par\blankline}

\def\date#1{\par\hbox to \hsize{\hfil #1\qquad}\par}

\def\ref{\par\noindent\hangindent 20pt}
\def\title#1\endtitle{\par\vfil\eject
     \par\vbox to 1.5in {}{\bf #1}\par\vskip 1.5in\nobreak}
\def\author#1\endauthor{\par\sc#1\par\blankline}

\def\sect#1\endsect{\par\vfil\eject{\bf #1}\par\vskip 12pt\nobreak}
\def\subsect#1\endsubsect{\vskip 14pt plus 50pt {\bf #1}\par
     \nobreak\blankline\nobreak}

\def\REFERENCES{\par\vfil\eject\vbox to 1in{}
     \ctrline{\bf REFERENCES}\par\vskip .5in }
% end of the file wbasict.tex

\pageno=1
\headline={\hss\tenrm\hss}
\singlespace
\nopagenumbers
\def\gsim{\;\lower4pt\hbox{${\buildrel\displaystyle >\over\sim}$}\;}
\def\lsim{\;\lower4pt\hbox{${\buildrel\displaystyle <\over\sim}$}\;}

\hbox{ }
\vskip 2.8cm
\ctrline{\bf DISTRIBUTIONS OF FOURIER MODES OF}
\ctrline{\bf COSMOLOGICAL DENSITY FIELDS}

\vskip 1.0cm
\ctrline{Zuhui Fan and J. M. Bardeen}

\vskip 1.0cm
\ctrline{\it Department of Physics, FM-15, University of Washington,
Seattle, WA 98195}

\vskip 2.0cm

\pageno=1
\headline={\hss\tenrm\folio\hss}
\doublespace
\nopagenumbers

\ctrline{\bf ABSTRACT}

We discuss the probability distributions of Fourier modes of
cosmological density fields
using the central limit theorem as it applies to weighted integrals
of random fields. It is shown that if the cosmological principle
holds in a certain sense, i.e., the universe approaches homogeneity and
isotropy sufficiently rapidly on very large scales, the one-point distribution
of each Fourier mode of the density field is Gaussian whether or not
the density field itself is Gaussian. Therefore, one-point
distributions of the power spectrum obtained from observational
data or from simulations are not a good test of whether the
density field is Gaussian.

PACS number(s): 98.80$_-$k, 98.80.BP, 98.60.Eg

\sect\ctrline{I. INTRODUCTION}\endsect

In the standard cold dark matter (CDM) model, primordial
fluctuations are generated during inflationary stage, and
the simple linear perturbation theory predicts Gaussian fluctuations [1].
On the other hand, some alternative inflationary theories [2] [3]
and topological defect mechanisms [4] [5] involve nonlinear processes
which generate non-Gaussian perturbations.
Thus it is interesting to test the consistency of a theory with
observations in this regard. Besides analyzing the spatial
distribution of galaxies [6] [7] and the pattern of microwave
background radiation anisotropies [8],
some research groups considered the one-point distribution of the power
spectrum
of the density field as a test of the Gaussianity [9] [10] [11] 12] [13].
It is found that for the QDOT IRAS redshift survey, the distribution is almost
exactly an exponential distribution which is consistent with the Gaussian
distribution for both the real and imaginary parts of a Fourier mode [13].
Suginohara and Suto [14] did simulations up to very nonlinear stages,
analyzed the one-point distribution for both the power spectrum and the phase
of a Fourier mode, and also got results consistent with Gaussian distributions.
On the other hand, we know that in highly
nonlinear stages the one-point density distribution in real space is
very non-Gaussian. This raises the question, namely, what can one
really conclude
from the statistics of the one-point distribution of the power spectrum?

The Fourier transform of the cosmological density field is basically
a weighted sum of the densities over the space. If the universe
consists of a large number of dense clumps (e.g., galaxies, clusters),
and those clumps are independent of each other, then by virtue of
the central limit theorem, the real and imaginary components of each
individual Fourier mode are Gaussian distributed although the density
distribution itself could be highly non-Gaussian.
Kaiser and Peacock [10] generated some test
catalogs, and analyzed the distributions of Fourier modes.
They found that even if only a few independent clumps existed
in a test catalog, the one-point probabilities of individual
Fourier modes would be well described by Gaussian distributions.

On the other hand, if there are some correlations on scales
larger than the typical size of a clump (as most of the cosmological
models predict), what is the one-point probability of
an individual Fourier mode?
In this paper, it is shown that if the random density field satisfies
certain mixing conditions, the central limit theorem
guarantees that the one-point distribution of an individual
Fourier component is Gaussian whether or not the field itself is Gaussian.

The paper is organized as following: In section II, we introduce
some concepts and discuss an application of the central limit theorem to
weighted integrals of a random field. In section III, we consider
the Fourier transform of a random field and see how it satisfies the conditions
in section II. Section IV contains further discussion and conclusions.
The appendix provides an approximate estimate of the mixing rate,
which is introduced and used in the section II, in terms of
the two-point joint probability.

\sect\ctrline{II. THE CENTRAL LIMIT THEOREM}\endsect

In this section, we discuss a central limit theorem for weighted
integrals of random scalar fields in terms of constraints on the field
moments, correlations and the weighting functions. The theorem
and its context is clearly discussed by
Ivanov and Leonenko [16]\footnote*{See the reference 16, Chapter I.}.

Let us first consider a random variable $\delta$. Without losing
generality, we assume the average $<\delta>=0$ and
$\delta \in [f_l, f_u]$, where $f_l \in R^1$ and $f_u \in R^1$.
Let $p(\delta)$ be the probability density function of $\delta$,
and a set $A_{\delta}$ be $\{\delta: f_l\le a_1\le \delta \le b_1 \le f_u\}$.
Then the probability that $\delta$ taking values on the set $A_{\delta}$
is
$$
P(A_{\delta})=\int_{a_1}^{b_1} p(\delta) d \delta \quad , \eqno (2.1)
$$
and the normalization condition is
$$
\int_{f_l}^{f_u} p(\delta) d \delta =1 \quad . \eqno (2.2)
$$

Now consider a random scalar field $\delta(\vec x)$
satisfying $<\delta(\vec x)>=0$, $\vec x \in R^3$.
We assume that
the probability density function
and the value range of $\delta(\vec x)$ are the same for all $\vec x \in R^3$.
Let $\Delta_1$ and $\Delta_2$ be two regions in $R^3$.
The distance between $\Delta_1$ and $\Delta_2$ is
defined as
$$
r=\min {|\vec x - \vec y|, \quad \vec x \in \Delta_1,
\quad \vec y \in \Delta_2} \quad .
$$
In order to apply the central limit theorem, we need to introduce
the concept of the mixing rate [17].
Let us denote positions on $\Delta_1$,
$\Delta_2$ by $\{\vec x_i, \hbox{} i=1,2,...\}$, $\{\vec y_j,
\hbox{} j=1,2,...\}$, respectively.
Let $A$ be a set generated by the random variables
$\{\delta(\vec x_i), \hbox{} i=1,2,...\}$ and
$$
a_i\le \delta(\vec x_i) \le b_i, \quad a_i \ge f_l,
\quad b_i \le f_u, \quad i=1,2,... \quad , \eqno(2.3)
$$
and let $B$ be generated by the random variables $\{\delta(\vec y_j),
\hbox{} j=1,2,...\}$ with the form
$$
c_j\le \delta(\vec y_j) \le e_j, \quad c_j \ge f_l,
\quad e_j \le f_u, \quad j=1,2,... \quad . \eqno(2.4)
$$
Then the probability $P(A)$, $P(B)$ and $P(AB)$ are
$$
P(A)=\{\Pi _{i=1,2,...}\int _{a_i}^{b_i} d \delta (\vec x_i)\}
\hbox{ } p[\delta(\vec x_1),\delta(\vec x_2),...] \quad , \eqno (2.5)
$$
$$
P(B)=\{\Pi _{j=1,2,...}\int _{c_j}^{e_j} d \delta (\vec y_j)\}
\hbox{ } p[\delta(\vec y_1),\delta(\vec y_2),...] \quad , \eqno (2.6)
$$
and
$$
P(AB)=\{\Pi _{i,j,=1,2,...}\int _{a_i}^{b_i}d \delta (\vec x_i)
\int _{c_j}^{e_j} d \delta (\vec y_j)\}
\hbox{ }p[\delta(\vec x_1),\delta(\vec x_2),...,
\delta(\vec y_1),\delta(\vec y_2),...] \quad .
\eqno (2.7)
$$
respectively, where $p[\delta(\vec x_1),\delta(\vec x_2),...]$
is the joint probability density function of $\{\delta(\vec x_i),
\hbox{} i=1,2,...\}$, $p[\delta(\vec y_1),\delta(\vec y_2),...]$
is the joint probability density function of $\{\delta(\vec y_j),
\hbox{} j=1,2,...\}$, and $p[\delta(\vec x_1),\delta(\vec x_2),...,
\delta(\vec y_1),\delta(\vec y_2),...]$ is the joint probability
density function of $\{\delta(\vec x_i),\delta(\vec y_j),
\hbox{} i=1,2,..., \hbox{} j=1,2,...\}$. The dependence of the random
variables defined on $\Delta_1$ and $\Delta_2$ is characterized
by the Rosenblatt dependence rate [17] which is defined as
$$
\alpha (\Delta_1, \Delta_2)=\max |P(AB)-P(A)P(B)| \quad ,
\eqno (2.8)
$$
where the maximum is taken over all possible values of
$\{ a_i, b_i \in [f_l,f_u], \quad c_j, e_j \in [f_l,f_u],
\quad i=1,2,..., \quad j=1,2,...\}$.
The mixing rate $\alpha(r_0)$ is defined as
$$
\alpha(r_0)=\max \alpha (\Delta_1, \Delta_2) \quad ,
\eqno (2.9)
$$
where the maximum is taken over all pairs of $\Delta_1$ and $\Delta_2$
lying at a distance at least $r_0$.
It is clear that in general the mixing
rate $\alpha(r_0)$ is not only related to the two-point
but also to all higher order correlations.

With these definitions, we now discuss the central limit theorem,
and introduce a column vector $s_V(\vec \theta)=[s^{(1)}_V(\vec \theta),
...,s^{(r)}_V(\vec \theta)]^t$, where $t$ stands for the transpose, with
components defined by
$$
s^{(i)}_V(\vec \theta)=\int_V g^{(i)}_V(\vec x,\vec \theta)
\delta(\vec x) d\vec x   \quad , \eqno(2.10)
$$
where $g_V(\vec x,\vec \theta)=[g^{(1)}_V(\vec x,\vec \theta),...,
g^{(r)}_V(\vec x,\vec \theta)]^t$ is a column vector of functions,
and $\vec \theta$ $\in \Theta$. In the situation we are going to
discuss in the next section, $\Theta = R^3$.

We assume that the following conditions are satisfied:

(i). As $V \rightarrow \infty$, there exists a function
${\cal \sigma}^2(\vec \theta)$, taking on values in a set
of positive definite $r\times r$ matrices such that
$$
\lim_{V \to \infty}[<s_V(\vec \theta)s^t_V(\vec \theta)>
-{\cal \sigma}^2(\vec \theta)] =0  \eqno(2.11)
 $$
uniformly in $\vec \theta \in \Theta$.

(ii). For any $i \in \{1,...,r\}$, the function $g^{(i)}_V
(\vec x,\vec \theta)$ satisfies the condition:

For $V \rightarrow \infty$, there exists a constant $k^{(i)}_3 >0$
(possibly depending on $i$) such that
$$
\max_{\vec \theta \in \Theta, \vec x \in V}
|g^{(i)}_V(\vec x,\vec \theta)|\le k^{(i)}_3/
\sqrt {V} \quad .
\eqno(2.12)
$$

(iii). There exists a constant $\gamma > 0$ such that
$$
<|\delta(\vec x)|^{2+\gamma}> \le k_1 < \infty, \quad \vec x\in R^3
\quad . \eqno (2.13)
$$

(iv). The mixing rate $\alpha(r_0)$ of the random field
$\delta(\vec x)$ satisfies
$$
\alpha(r_0) \le k_{2}r_0^{-3-\epsilon} \quad , \quad \epsilon \gamma > 6
\quad ,
\eqno(2.14)
$$
where $k_2$ is a constant.

Then as $V \rightarrow \infty$, the random vector $s_V
(\vec \theta)$ has, uniformly in $\Theta$, an asymptotically
multidimensional normal distribution $N_r(0,{\cal \sigma}^2(\vec \theta))$
with zero mean vector and covariance matrix
${\cal \sigma}^2(\vec \theta)$\footnote*{See the reference 16, p41,
{\bf Theorem 1.7.5}}.

In the next section, we will discuss the Fourier transform of
a cosmological density field using this central limit theorem.

\sect\ctrline{III. FOURIER TRANSFORM OF A RANDOM FIELD}\endsect

The Fourier transform of a random field $\delta(\vec x)$ in
a 3-dimensional space is defined as
$$
\delta(\vec k)=\lim_{V \to \infty} {1 \over \sqrt V} \int_V
\hbox{}\exp (i\vec k \cdot {\vec x}) \delta(\vec x) d^3\vec x
\quad . \eqno (3.1)
$$
Here we assume $\delta (\vec x)$ is a homogeneous, isotropic and
real random field, which satisfies $<\delta (\vec x)>=0$.

The Fourier transform $\delta(\vec k)$ can be written
as a sum of the real part $A(\vec k)$
and the imaginary part $iB(\vec k)$, where $A(\vec k)$ and $B(\vec k)$
are defined as
$$
A(\vec k)=\lim_{V \to \infty}A_{V}(\vec k) \quad , \eqno (3.2)
$$

and
$$
B(\vec k)=\lim_{V \to \infty}B_{V}(\vec k) \quad , \eqno (3.3)
$$
respectively, where
$$
A_V={1 \over \sqrt V} \int_V \hbox{} \cos (\vec k \cdot {\vec x})
\delta (\vec x)
\quad ,
$$

and
$$
B_V={1 \over \sqrt V} \int_V \hbox{} \sin (\vec k \cdot {\vec x})
\delta (\vec x)
\quad .
$$

We consider the vector $s_{V}(\vec k)=[A_{V}(\vec k), B_{V}(\vec k)]$.
In terms of the notations of section II, we have
$r=2$ and
$$
g^{(1)}_{V}(\vec x, \vec k) = {1\over \sqrt V} \cos (\vec k \cdot
{\vec x})  \quad , \eqno(3.4)
$$

$$
g^{(2)}_{V}(\vec x, \vec k) = {1\over \sqrt V} \sin (\vec k \cdot
{\vec x}) \quad . \eqno(3.5)
$$

We discuss the conditions of the central limit
theorem in the section II one by one.

(i). To evaluate the covariance matrix ${\cal \sigma}^2(\vec k)$,
three elements should be considered:
$${\cal \sigma}^2_{11}(\vec k)=\lim_{V \to \infty}<A_V(\vec k)A_V(\vec k)>
\quad , $$
$${\cal \sigma}^2_{22}(\vec k)=\lim_{V \to \infty}<B_V(\vec k)B_V(\vec k)>
\quad , $$
and
$${\cal \sigma}^2_{12}(\vec k)=\lim_{V \to \infty}<A_V(\vec k)B_V(\vec k)>
\quad . $$

First let us consider ${\cal \sigma}^2_{11}$.
$$
{\cal \sigma}^2_{11}(\vec k)=\lim_{V \to \infty}
\int_{V}\int_{V} d^3{\vec x}_1 d^3{\vec x}_2
{1\over \sqrt V} \cos (\vec k \cdot {{\vec x}_1})
{1\over \sqrt V} \cos (\vec k \cdot {{\vec x}_2})
<\delta (\vec x_1) \delta (\vec x_2)>  \quad . \eqno (3.6)
$$
Because $\delta (\vec x)$ is homogeneous, isotropic and
$<\delta (\vec x_1)>=0$, $<\delta (\vec x_2)>=0$, the correlation
$<\delta (\vec x_1) \delta (\vec x_2)>=\xi (|\vec x_1-\vec x_2|)$
only depends on $|\vec x_1 -\vec x_2|$. One can readily show that
$$
{\cal \sigma}^2_{11}(\vec k)={1\over 2} \lim_{V \to \infty}
\int_{V} d^3{\vec x} \cos(\vec k \cdot {\vec x}) \xi (|\vec x|)
\quad . $$

We know that the power spectrum averaged over the ensemble
of realizations is related to the correlation function
$\xi (\vec x)$ by
$$
<P(\vec k)>=\lim_{V \to \infty} \int_{V} d^3{\vec x}
\exp(i\vec k \cdot \vec x) \xi (\vec x) \quad .
$$
Since $\delta (\vec x)$ is homogeneous, isotropic and real,
it follows that

\noindent
$< P(\vec k)>=<P(-\vec k)>=<P(k)>$
and $\xi (\vec x)=\xi (|\vec x|)$.
Thus
$$
\lim_{V \to \infty}
\int_{V} d^3{\vec x} \cos(\vec k \cdot {\vec x}) \xi (|\vec x|)
=<P(k)> \quad ,
$$
and
$$
\lim_{V \to \infty}
\int_{V} d^3{\vec x} \sin (\vec k \cdot {\vec x}) \xi (|\vec x|)=0
\quad . $$
Then
$$
{\cal \sigma}^2_{11}(\vec k)={1\over 2}<P(k)> \quad . \eqno (3.7)
$$

Similarly one can show that
$$
{\cal \sigma}^2_{22}(\vec k)={1\over 2}<P(k)> \quad , \eqno (3.8)
$$
and
$$
{\cal \sigma}^2_{12}(\vec k)=0 \quad . \eqno (3.9)
$$
For any reasonable cosmological density field,
the average power spectrum

\noindent
$<P(k)>$ exists.
Therefore the condition for the existence of
the covariance matrix ${\cal \sigma}^2(\vec k)$ is satisfied.

(ii). From (3.4), (3.5), $|\cos(\vec k \cdot {\vec x})|\le 1$,
and $|\sin(\vec k \cdot {\vec x})|\le 1$, we can always choose
$k_3= 1$ so that condition (ii) in section II is
satisfied.

Condition (iii) of section II constrains the statistical properties
of the field $\delta (\vec x)$.
The existence of moments higher than the second
is not a problem for most models of cosmological density field.

Condition (iv) of section II concerns the behavior of the mixing
rate $\alpha (r_0)$.
In general this mixing rate is related to
all order of correlations and thus is difficult to be calculated
precisely. Under certain approximations, however, we can estimate the
behavior of $\alpha (r_0)$. In the Appendix,
we make such an approximation:
Between two well separated regions $\Delta_1$ and $\Delta_2$,
only two-point joint probabilities are considered,
while within each region full correlations are taken into account.
We argue in the Appendix that under plausible
assumptions, the reduced two-point joint probability
of $\delta_i$ at $\vec x_i$ and $\delta_j$ at $\vec y_j$ can be
written as
$$
\tilde {p}(\delta_i, \delta_j, |\vec x_i-\vec y_j|)=\sum_{n=n_0}^\infty
A_n(\delta_i,\delta_j) F_{ij}^n(|\vec x_i-\vec y_j|),
\eqno (3.10)
$$
where $F_{ij}(|\vec x_i-\vec y_j|)\lsls 1$ when $|\vec x_i-\vec y_j|$
is very large. $A_n(\delta_i,\delta_j)$ is independent of
$F_{ij}(|\vec x_i-\vec y_j|)$ and $n_0$ is the power index
such that $A_n$ vanishes for $n<n_0$. Here $n_0\ge 1$.

Under these approximations
$$
\alpha(r_0)\sim C_0^{\prime}\max_{|\vec x_i -\vec y_j|\ge r_0}
|F_{ij}^{n_0} (|\vec x_i-\vec y_j|)|, \eqno (3.11)
$$
where $C_0^{\prime}$ is a finite constant, and
the maximum is taken over all pairs of ($\vec x_i, \vec y_j$)
with $|\vec x_i -\vec y_j|\ge r_0$.
The condition (iv) of section II requires that
$$
\alpha(r_0) \le k_{2} r_0^{-3-\epsilon} \quad ,
\quad \epsilon \gamma > 6 \quad .
\eqno(3.12)
$$
Then in order the inequality (3.12)
to be satisfied, we need
$$
\max_{|\vec x_i -\vec y_j|\ge r_0}
|F_{ij}^{n_0} (|\vec x_i-\vec y_j|)|\le C_{14} r_0^{-3-\epsilon}
\quad , \eqno (3.13)
$$
where $C_{14}$ is a finite constant. As discussed in the Appendix,
for a non-Gaussian field which is a local functional of a Gaussian
field, $F_{ij}(|\vec x_i-\vec y_j|)$ is the two-point correlation
function of the Gaussian field. The power index $n_0$ in (3.10) depends on the
specific form of the non-Gaussian field. For example, for a
field $\Phi=\phi^2-<\phi^2>$, where $\phi$ is a Gaussian field
and $<\phi^2>$ is the average of $\phi^2$, the power index $n_0$ is $2$.
If the two-point correlation function of a Gaussian field
satisfies condition (3.13) (in this case $n_0=1$), all non-Gaussian fields
which are local functionals of the Gaussian field satisfy
the mixing rate condition.

Reduced higher order joint probabilities are related to higher order
reduced correlations. Observations indicate that
higher order of reduced correlations drop faster
than the two-point correlation function [15] [18] [19]. Thus
it is a reasonable approximation for a cosmological density field
to ignore reduced higher order joint probabilities.

As a case where the mixing rate condition
is violated, we consider a random field $\delta (\vec x)$
which is a convolution of two Gaussian fields
$\phi_1$ and $\phi_2$
$$
\delta (\vec x)=\int d^3\vec x_1 \phi_1(\vec x_1) \phi_2 (\vec x-\vec x_1)
\quad . \eqno (3.14)
$$
Clearly, for each $\vec x \in R^3$, $\delta (\vec x)$ is related
to the fields $\phi_1$ and $\phi_2$ in the whole space. Thus
the probabilities of the field $\delta (\vec x)$ is always
related to the joint probabilities of all $\phi_1$ and all $\phi_2$
in the whole space. It is not surprising that some
complicated correlations for the random field $\delta (\vec x)$ exist
and the mixing rate condition is not satisfied. In fact,
we know that the Fourier transform of $\delta (\vec x)$ is the
product of the Fourier transform of $\phi_1$ and $\phi_2$
$$
\delta (\vec k)=\phi_1 (\vec k) \phi_2 (\vec k) \quad , \eqno (3.15).
$$
so $\delta (\vec k)$ is not Gaussian distributed.

If a field $\delta (\vec x)$ satisfies the conditions
of the central limit theorem, then the the joint
distribution of the real and imaginary
parts of its Fourier transform, namely, $A(\vec k)$ and $B(\vec k)$,
is simply
$$
p(A,B)dAdB=\bigg({1\over  {\pi <P(k)>}}\bigg) \exp \bigg[-({A^2+B^2 \over
<P(k)>^2})\bigg] dAdB \quad .   \eqno (3.16)
$$

If we change the variables from $A(\vec k)$ and $B(\vec k)$ to
$P(\vec k)$ and $\theta_{\vec k}$, where $\theta_{\vec k}$
is the phase of the $\vec k$ mode, we get
$$
p\bigg[P(\vec k),\theta_{\vec k}\bigg]=\exp \bigg[-{P(\vec k)\over <P(k)>}
\bigg] dPd\theta_{\vec k} \quad , \eqno (3.17)
$$
that is, $P(\vec k)$ is exponentially distributed, and $\theta_
{\vec k}$ is uniformly distributed.

\sect\ctrline{IV. DISCUSSION}\endsect

In this paper we showed that as long as
the mixing rate of the density field drops fast enough at
large distances, the central limit theorem guarantees
normal distributions for the real and imaginary parts
of a Fourier mode of the field.

Reduced higher order joint probabilities are related
to the reduced higher order correlations. Observational
evidences show that on large scales, these correlations do drop faster
than the two-point correlation function does [15] [18] [19]. Therefore
for the mixing rate, it is reasonable to only consider the two-point joint
probabilites between two well separated regions.
Under such an approximation,
we estimate the mixing rate. If the reduced two-point joint
probability of $\delta_i$ at $\vec x_i$ and $\delta_j$ at
$\vec y_j$ depends on a function $F_{ij}(|\vec x_i-\vec y_j|)$
in such a way
$$
\tilde {p}(\delta_i, \delta_j, |\vec x_i-\vec y_j|)=\sum_{n=n_0}^\infty
A_n(\delta_i,\delta_j) F_{ij}^n(|\vec x_i-\vec y_j|),
$$
where $|F_{ij}^n(|\vec x_i-\vec y_j|)|\lsls 1$ when
$|\vec x_i-\vec y_j|$ is very large, and $A_n(\delta_i,\delta_j)$
is independent of
$F_{ij}^n(|\vec x_i-\vec y_j|)$, the mixing rate has the same
behavior as
$$
C_0^{\prime}\max_{|\vec x_i -\vec y_j|\ge r_0}
|F_{ij}^{n_0} (|\vec x_i-\vec y_j|)|,
$$
where $C_0^{\prime}$ is a finite constant. Thus if
$$
\max_{|\vec x_i -\vec y_j|\ge r_0}|F_{ij}^{n_0} (|\vec x_i-\vec y_j|)|
\le C_{14}r_0^{-3-\epsilon}
$$
at large $r_0$, where $\epsilon>0$,
the mixing rate condition is satisfied.
The condition on
$\max_{|\vec x_i -\vec y_j|\ge r_0}|F_{ij}^{n_0} (|\vec x_i-\vec y_j|)|$
is equivalent to requiring that
$$
\max_{|\vec x_i -\vec y_j|\ge r_0}
\tilde {p}(\delta_i, \delta_j, |\vec x_i-\vec y_j|)
\le C_{15}r_0^{-3-\epsilon}.
$$
This is a sensible requirement, and
is easily satisfied by a cosmological density field.

Statistical properties of the large-scale structure in the universe
provide important clues for understanding the universe.
Theoretically, to analyze statistical properties
of a random field, one must generate many different
realizations of the field and calculate the
statistical quantities over these realizations. Practically,
since there is only one universe,
we have to use the `` ergodic'' theorem in the position
space to deal with cosmological
density field. Therefore
it is important to have what is in some sense `` a fair
sample ''. The basic requirement for ``a fair sample'' is
that quantities within the sample should represent the
quantities of the whole universe without being strongly biased by the
correlations of the quantities within the sample.
If the universe is statistically homogeneous and isotropic at
very large scales, all correlations within ``a fair sample''
should have no significant effect on the statistical quantities
and can be ignored.
For such samples, the Fourier modes of the density field
are Gaussian distributed whether or not the density field itself is Gaussian.
Since in reality one can never be
certain whether or not we really have a fair sample,
we must be cautious to interpret the statistical
results on the distributions of the Fourier modes.
If non-Gaussian distributions of individual Fourier modes
are found for a set of samples and this non-Gaussianity
disappears when larger samples are taken, it is most likely
that the original sample is not a fair sample,
rather than that there is a genuine non-Gaussian behavior.
If the non-Gaussianity persists when we go to larger and larger samples,
then one might begin to believe that there is some physics in it.

One class of non-Gaussian models are topological defect seeded
models, e.g., cosmic strings [5], or textures [4].
In those models, the density perturbations are due to
randomly distributed seeds, and each seed has a density profile.
If the density profiles of
all seeds are the same, and are denoted by $f(\vec x -\vec x_i)$,
where $\vec x_i$ is the position of a seed,
the density field can be written as
$$
\delta(\vec x)=\sum_{i=1}^N f(\vec x -\vec x_i), \eqno(4.1)
$$
where the sum is over all the seeds.
We can write (4.1) as
$$
\delta(\vec x)=\int d\vec x^{\prime}
g(\vec x^{\prime})f(\vec x -\vec x^{\prime}),
\eqno(4.2)
$$
where
$$
g(\vec x^{\prime})=\sum_{i=1}^N \delta^{(3)}(\vec x^{\prime}-\vec x_i),
\eqno(4.3)
$$
and $\delta^{(3)}(\vec x^{\prime}-\vec x_i)$ is the three-dimensional
Dirac $\delta$-function.
Then the Fourier transform of $\delta(\vec x)$ is
$$
F_{\delta}(\vec k)=F_{g}(\vec k)F_{f}(\vec k), \eqno(4.4)
$$
where $F_{g}(\vec k)$ is the Fourier transform of the function $g$,
and $F_{f}(\vec k)$ is the Fourier transform $f$.
Since $f$ is not a random function, $F_{f}(\vec k)$ is not
random. The distribution of $F_{\delta}(\vec k)$ depends on
the statistics of $F_{g}(\vec k)$.
We have
$$
F_{g}(\vec k)={1\over \sqrt V}\sum_{i=1}^N \exp(i\vec k \cdot \vec x_i).
\eqno(4.5)
$$
If the seeds are uncorrelated and $N\rightarrow \infty$,
$F_{g}(\vec k)$ is Gaussian distributed and so is
$F_{\delta}(\vec k)$ by the central limit theorem for
the discrete summation[20].
Certainly, if within an observational region
only one or two seeds exist, one may get some
non-Gaussian signal in the distributions of the
Fourier transform [20]. But in that case, one needs to go
to larger samples in order to get statistically solid
conclusions.

In principle, the appropriate way to analyze the distributions
of the Fourier modes (power spectrum) is to generate a large
number of realizations of the cosmological density field.
For each realization, the Fourier transform is made, and
the average power spectrum $<P(k)>$ for each $k$
is obtained by averaging over all realizations.
Then the distribution of $P(k)/<P(k)>$ for each $k$
is analyzed. Practically, since we only have one universe,
and can do limited number of simulations, some kind of
``ergodic'' assumptions are always used when analyzing
observational or simulation data.
For example, in [12] and [13] the distributions of the power
spectrum in $k$-space are analyzed, {\it i.e.},
to cumulate the numbers of $k$ which have the same value of $P/<P>$,
where $<P>$ is the average power based on
some physical considerations. Then the distributions
of those numbers against $P/<P>$ are analyzed.
In these analyses, we should be very careful to use an
appropriate $<P>$. For example, as pointed out in [13],
because of the small scale clustering,
the average $<P(k)>$ is larger than the purely random
walk estimation for large $k$. If one uses $<P>$ estimated by ignoring
the clustering effect, one may get a power spectrum
distribution which is different from the exponential
distribution expected from Gaussian distributions
of Fourier modes. This difference is caused by
using an inappropriate average power spectrum, rather
than the existence of some genuine non-Gaussianity.

By the above discussion, we see that the one-point
distributions of Fourier modes are not good
statistical tests of the Gaussianity of the density field.
One possible way to distinguish Gaussian and non-Gaussian statistics
is to go to high order correlations. For example, we can consider
the two-point correlation of the power spectra $F(\vec k, \vec k')
=<P(\vec k)P(\vec k')> $, which is related
to the four-point correlation of Fourier components [13].
For a Gaussian random field $\phi (\vec x)$,
$$\eqalign{
F(\vec k, \vec k')=&<P(\vec k)> <P(\vec k')>
\cr & + \delta ^{(3)}
(\vec k +\vec k')<P(\vec k)> <P(\vec k)>
\cr &+\delta ^{(3)}(\vec k -\vec k')<P(\vec k)><P(\vec k)>
\cr }
\eqno (4.1)
$$

If $\vec k + \vec k'\neq 0$ and $\vec k - \vec k'\neq 0$, then
$$
F(\vec k, \vec k')=<P(\vec k)> <P(\vec k')>
\eqno (4.2)
$$
{\it i.e.}, for a Gaussian field, the correlation of the
power spectrum for different $k$-mode is simply the product of
the power spectrum of each mode.

For a non-Gaussian field, it is expected that there are some
additional terms in the correlation of the power spectrum.
To illustrate this point, as an example, we consider a field $\zeta (\vec x)$
which is the square of a Gaussian random field  $\phi (\vec x)$.
$$
\zeta (\vec x)=\phi ^2 (\vec x) \eqno (4.3)
$$
Detailed calculations show that if $\vec k \neq -\vec k'$
and $\vec k \neq \vec k'$,
$$\eqalign{
F(\vec k, \vec k')=& <P(\vec k)> <P(\vec k')>
\cr & + 8\int {d^3 \vec k_1 \over (2 \pi)^3}
<|\phi (\vec k_1)|^2 |\phi (\vec k -\vec k_1)|^2
|\phi (\vec k_1+\vec k')|^2 |\phi (\vec k -\vec k'-\vec k_1)|^2>
\cr & + 8\int {d^3 \vec k_1 \over (2 \pi)^3}
<|\phi (\vec k_1)|^2 |\phi (\vec k -\vec k_1)|^2
|\phi (\vec k_1-\vec k')|^2 |\phi (\vec k +\vec k'-\vec k_1)|^2>
\cr }
\eqno (4.4)
$$
It can be seen that there are some extra terms added up to
$<P(\vec k)> <P(\vec k')>$. Generally these extra terms depend on the
specific form of the non-Gaussian field. Therefore at
this level, it is possible to distinguish Gaussian and non-Gaussian fields.

\vskip 1.0cm
\ctrline{ACKNOWLEDGEMENTS}

We are very grateful for helpful discussions with P. Guttorp and
N. Kaiser. And we are very thankful to N. Kaiser for providing us their
preprint. One of us (JMB) acknowledges useful discussions with other
participants at an Aspen Center for Physics Workshop. This work was
supported in part by the Department of Energy under Grant No.
DE-FG06-91ER40614.

\REFERENCES
\ref
[1] J. M. Bardeen, P. J. Steinhardt,  and M. S. Turner,
Phys. Rev. D. {\bf 28}, 679 (1983)

\ref
[2] D. S. Salopek, J. R. Bond, and J. M. Bardeen,
Phys. Rev. D. {\bf 40}, 6 (1989)

\ref
[3] L. Kofman, G. R. Blumenthal, H. Hodges, and J. R. Primack,
{\it Large Scale Structures and Peculiar Motions in the Universe}
(ed. David W. Latham and L. A. Nicolaci da Costa, Astronomical
Society of the Pacific), P339 (1991)

\ref
[4] N. Turok, Phys. Rev. Lett. {\bf 63} 2625 (1989)

\ref
[5] A. Vilenkin, Phys. Rev. Lett., {\bf 46}, 1169 (1981)

\ref
[6] W. Saunders, C. S. Frenk, M. Rowan-Robinson, G. Efstathiou,
A. Lawrence, N. Kaiser, R. S. Ellis, J. Crawford, X. Y. Xia,
and I. Parry, Nature, {\bf 349}, 32 (1991)

\ref
[7] G. Efstathiou, N. Kaiser, W. Saunders, A. Lawrence,
M. Rowan-Robinson, R. S. Ellis, and C. S. Frenk,
MNRAS, {\bf 247}, 10p (1990)

\ref
[8] D. Coulson, P. Ferreira, P. Graham, and N. Turok,
Nature, {\bf 368}, 27 (1994)

\ref
[9] T. J. Broadhurst, R. S. Ellis, D. C. Koo, and
A. S. Szalay, Nature {\bf 343}, 726 (1990).

\ref
[10] N. Kaiser, and J. A. Peacock, ApJ {\bf 379}, 482 (1991).

\ref
[11] D. Tytler, J. Sandoval, and X. M. Fan, ApJ {\bf 405}, 57 (1993).

\ref
[12] A. S. Szalay, R. S. Ellis, D. C. Koo, and T. J. Broadhurst,
{\it After the First Three Minutes} (AIP Conference Proceedings 222,
ed. Holt, Bennett and Trimble, AIP, New York 1990).

\ref
[13] H. A. Feldman, N. Kaiser, and J. A. Peacock,  ApJ, {\bf 426},
23 (1994)

\ref
[14] T. Suginohara, and Y. Suto, ApJ {\bf 371}, 470 (1991).

\ref
[15] P. J. E. Peebles,
{\it The Large Scale Structure of the Universe}
(Princeton University Press, Princeton, USA, 1980).

\ref
[16] A. V. Ivanov, and N. N. Leonenko,
{\it Statistical Analysis of
Random Fields} (Kluwer Academic Publishers, 1989).

\ref
[17] M. Rosenblatt, Proc. Natl. Acad. Sci, U.S.A. {\bf 42}, 43 (1956)

\ref
[18] G. Toth, J. Hollos, and A. S. Szalay, ApJ, {\bf 344}, 75 (1989)

\ref
[19] J. R. Gott III, B. Gao, and C. Park, ApJ, {\bf 383}, 90 (1991)

\ref
[20] L. Perivolaropoulos, MNRAS, {\bf 267}, 529 (1994)

\sect\ctrline{Appendix}\endsect
\ctrline{\bf Estimation of the Mixing Rate}

In this appendix, we discuss the approximate behavior of the mixing
rate.

Consider $|P(AB)-P(A)P(B)|$ as discussed in section II.
If the random variables in $\Delta_1$ are
completely independent of the random variables
in $\Delta_2$, then $P(AB)=P(A)P(B)$,
and $|P(AB)-P(A)P(B)|=0$ as expected.
If the two sets of random variables are weakly correlated,
we could only consider
the first order corrections to $P(AB)$ in addition to $P(A)P(B)$.
We consider a statistically homogeneous and isotropic random field.
For two random variables $\delta_i$ at $\vec x_i$ and $\delta_j$ at
$\vec y_j$,
their joint probability is $p(\delta_i, \delta_j, |\vec x_i-\vec y_j|)$.
Then
$$
\tilde{p}(\delta_i, \delta_j, |\vec x_i-\vec y_j|)
=p(\delta_i, \delta_j, |\vec x_i-\vec y_j|)-p(\delta_i)p(\delta_j)
\eqno (A.1)
$$
is the reduced two-point joint probability which represents
the contribution to $p(\delta_i, \delta_j, |\vec x_i-\vec y_j|)$
due to the correlation between $\delta_i$
and $\delta_j$.
The first order approximation of $P(AB)-P(A)P(B)$
means that between two well separated regions, only reduced two-point
joint probabilities are considered. Let us consider
the contribution from the correlation between
$\delta_i\in \Delta_1$ and $\delta_j\in \Delta_2$. After
$\delta_i\in \Delta_1$ having been picked out,
since $\delta_i$ is correlated with
the rest of the variables within $\Delta_1$, the joint probability
for the rest of the random variables within $\Delta_1$ should be
the conditional joint probability with $\delta_i$ fixed, which is
denoted by $p(\delta(\vec x_1),..., \delta(\vec x_{i-1}),
\delta(\vec x_{i+1}),...|\delta(\vec x_i))$. Similarly, the joint
probability for the rest of the variables within $\Delta_2$ should be
$p(\delta(\vec y_1),..., \delta(\vec y_{j-1}),
\delta(\vec y_{j+1}),...|\delta (\vec y_j))$, the conditional probability
with $\delta_j$ fixed. Therefore the contribution to $P(AB)-P(A)P(B)$
from the correlation between $\delta_i$ and $\delta_j$ has the form
$$\eqalign{
P_{ij}=&
\bigg\lbrace  \int_{a_{i}}^{b_{i}}
d \delta(\vec x_i) \int_{c_{j}}^{e_{j}} d \delta(\vec y_j)
\tilde {p}(\delta_i, \delta_j, |\vec x_i-\vec y_j|)
\cr &
\Pi _{p_i=1,...,i-1,i+1,..., q_j=1,...j-1,j+1,...}
\int_{a_{p_i}}^{b_{p_i}} d \delta(\vec x_{p_i})
\int_{c_{q_j}}^{e_{q_j}} d \delta(\vec y_{q_j})
\cr &
p(\delta(\vec x_1),..., \delta(\vec x_{i-1}), \delta(\vec x_{i+1}),...|
\delta(\vec x_i)) \cr &
\times p(\delta(\vec y_1),..., \delta(\vec y_{j-1}), \delta(\vec y_{j+1}),...|
\delta(\vec y_j))
\bigg\}. \cr} \eqno (A.2)
$$
The first order correction to $P(AB)-P(A)P(B)$ should include contributions
from all pairs of random variables between $\Delta_1$ and $\Delta_2$.
When $r_0$ is sufficiently large, approximately we have
$$
P(AB)-P(A)P(B)\sim \sum _{i,j=1,2,...} P_{ij},
\eqno (A.3)
$$
where $P_{ij}$ has the form (A.2)

Since we are interested in the behavior of the mixing rate with large
separation between the two regions,
we would like to separate the correlations
between $\delta_i$ and $\delta_j$, which depends on
$|\vec x_i-\vec y_j|$ for a statistically homogeneous
and isotropic random field, from the statistics for
each of the random variables. First, let us consider
a non-Gaussian field $\delta(\vec x)$ which is a local
functional of a Gaussian field $\phi(\vec x)$, {\it i.e.},
$\delta(\vec x)=\delta[\phi(\vec x)]$. The joint probability
between $\delta(\vec x_i)$ and $\delta(\vec y_j)$ is related to
the joint probability of $\phi(\vec x_i)$ and $\phi(\vec y_j)$.
We assume that for a fixed value of $\delta(\vec x)$, there are
$m$ values of $\phi^{(k)}(\vec x),(k=1,...,m)$ corresponding to it,
it then follows
$$\eqalign{
p[\delta(\vec x_i),\delta(\vec y_j)] &
=\sum_{k_i=1,k_j=1}^{m}p[\phi^{(k_i)}
(\vec x_i),\phi^{(k_j)}(\vec y_j)]
\cr &
\bigg\{{1\over |d\delta[\phi(\vec x_i)]/ d[\phi(\vec x_i)]|_{k_i}}\bigg\}
\bigg\{{1\over |d\delta[\phi(\vec y_j)]/ d[\phi(\vec y_j)]|_{k_j}}\bigg\rbrace,
\cr }
\eqno (A.4)
$$
where the factor
$\{1/ |d\delta[\phi(\vec x_i)]/ d[\phi(\vec x_i)]|_{k_i}\}
\{1/ |d\delta[\phi(\vec y_j)]/ d[\phi(\vec y_j)]|_{k_j}\}$
comes from the Jacobi transformation taking value at
$\phi^{(k_i)}(\vec x_i)$ and $\phi^{(k_j)}(\vec y_j)$.
It is well known that the two-point joint probability
$p[\phi(\vec x_i),\phi(\vec y_j)]$ of a Gaussian
field $\phi$ with $<\phi^2>=1$ can be expanded as
$$
p[\phi(\vec x_i),\phi(\vec y_j)]=\sum_{n=0}^{\infty}
{\xi_{\phi}^{n}(|\vec x_i-\vec y_j|)\over n!}
\Phi^{(n+1)}[\phi(\vec x_i)]\Phi^{(n+1)}[\phi(\vec y_j)],
\eqno (A.5)
$$
where $\xi_{\phi}(|\vec x_i-\vec y_j|)$ is the two-point
correlation function of the Gaussian field $\phi (\vec x)$, and
$$
\Phi^{(n+1)}[\phi(\vec x_i)]={1\over 2\pi}\exp[-\phi^2(\vec x_i)/2]
H_{n}[\phi(\vec x_i)],
\eqno(A.6)
$$
where $H_{n}$ is the Chebyshev-Hermite polynomials
\footnote*{See the reference 16, Page 55}.
Then (A.4) can be written as
$$
p[\delta(\vec x_i),\delta(\vec y_j)]=
\sum_{n=0}^{\infty}\xi_{\phi}^{n}(|\vec x_i-\vec y_j|)
A_{n}[\delta(\vec x_i),\delta(\vec y_j)],
\eqno (A.7)
$$
where
$$\eqalign{
A_{n}[\delta(\vec x_i),\delta(\vec y_j)]=& {1\over n!}
\sum_{k_i=1,k_j=1}^{m}\bigg\{({1\over |d\delta[\phi(\vec x_i)]/
d[\phi(\vec x_i)]|_{k_i}})\Phi^{(n+1)}[\phi^{(k_i)}(\vec x_i)]
\cr &
+({1\over |d\delta[\phi(\vec y_j)]/ d[\phi(\vec y_j)]|_{k_j}})
\Phi^{(n+1)}[\phi^{(k_j)}(\vec y_j)]\bigg\}\cr }.
\eqno (A.8)
$$
The reduced two-point probability is
$$
\tilde p[\delta(\vec x_i),\delta(\vec y_j)]=
\sum_{n=1}^{\infty}\xi_{\phi}^{n}(|\vec x_i-\vec y_j|)
A_{n}[\delta(\vec x_i),\delta(\vec y_j)].
\eqno (A.9)
$$
For a specific non-Gaussian field, the index $n$ of the first
nonvanishing $A_{n}[\delta(\vec x_i),\delta(\vec y_j)]$ can be
greater than $1$. We can write (A.9) as
$$
\tilde p[\delta(\vec x_i),\delta(\vec y_j)]=
\sum_{n=n_0}^{\infty}\xi_{\phi}^{n}(|\vec x_i-\vec y_j|)
A_{n}[\delta(\vec x_i),\delta(\vec y_j)],
\eqno (A.10)
$$
where $n_0$ is the index such that all $A_n (n<n_0)$ vanish.
The dependence of $\tilde p[\delta(\vec x_i),\delta(\vec y_j)]$
on $|\vec x_i-\vec y_j|$  is thus shown explicitly in (A.10).

Based on the above considerations, we make several assumptions:

\noindent
(i). $\tilde {p}(\delta_i, \delta_j, |\vec x_i-\vec y_j|)$ depends
on a function $F_{ij}(|\vec x_i-\vec y_j|)$ and can be written as
$$
\tilde {p}(\delta_i, \delta_j, |\vec x_i-\vec y_j|)=\sum_{n=n_0}^\infty
A_n(\delta_i,\delta_j) F_{ij}^n(|\vec x_i-\vec y_j|),
\eqno (A.11)
$$
where $A_n(\delta_i,\delta_j)$ is independent of $F_{ij}(|\vec x_i-\vec y_j|)$.
When $F_{ij}(|\vec x_i-\vec y_j|)=0$, (A.11) implies that $\delta_i$
and $\delta_j$ are independent of each other and
$\tilde {p}(\delta_i, \delta_j, |\vec x_i-\vec y_j|)=0$.
For a non-Gaussian
field which is a local functional of a Gaussian field $\phi$,
$F_{ij}(|\vec x_i-\vec y_j|)=\xi_{\phi}(|\vec x_i-\vec y_j|)$
as shown above.
The random variables can have value ranges extending to
infinity, so it is necessary to keep all the terms in the expansion (A.11).

\noindent
(ii). When $|\vec x_i-\vec y_j|$ is large,
$F_{ij}(|\vec x_i-\vec y_j|)\lsls 1$.

By (A.11), $P_{ij}$ in (A.2) can be written as
$$
P_{ij}=\sum_{n=n_0}^\infty F_{ij}^n(|\vec x_i-\vec y_j|)
C_{n_{ij}} \eqno (A.12)
$$
where
$$\eqalign{
C_{n_{ij}}=&
\bigg \{ \int_{a_{i}}^{b_{i}}
d \delta(\vec x_i) \int_{c_{j}}^{e_{j}} d \delta(\vec y_j)
A_n(\delta_i,\delta_j)
\cr &
\Pi _{p_i=1,...,i-1,i+1,..., q_j=1,...j-1,j+1,...}
\int_{a_{p_i}}^{b_{p_i}} d \delta(\vec x_{p_i})
\int_{c_{q_j}}^{e_{q_j}} d \delta(\vec y_{q_j})
\cr &
p(\delta(\vec x_1),..., \delta(\vec x_{i-1}), \delta(\vec x_{i+1}),...|
\delta(\vec x_i)) \cr &
\times p[\delta(\vec y_1),..., \delta(\vec y_{j-1}), \delta(\vec y_{j+1}),...|
\delta(\vec y_j)]
\bigg \}, \cr} \eqno (A.13)
$$
is a function of $(\vec x_1,...,\vec x_i,...,a_1,b_1,...,a_i,b_i,...)$
and $(\vec y_1,...,\vec y_i,..., c_1,e_1,...,c_i,e_i,...)$.
Since $C_{n_{ij}}$ is finite,
by the assumption (ii), we only consider $F_{ij}^{n_0}(|\vec x_i-\vec y_j|)$
and ignore all the other higher order of $F_{ij}(|\vec x_i-\vec y_j|)$.
Then
$$
P(AB)-P(A)P(B)\sim \sum_{i,j}F_{ij}^{n_0}(|\vec x_i-\vec y_j|)
C_{n_{0_{ij}}}
\eqno (A.14)
$$
For a specific random field, $F_{ij}$ may have a certain
value range. It is clear that $F_{ij}=0$ must be
an allowable value.

We also have the following relation:
$$
|P(AB)-P(A)P(B)|\sim |\sum_{i,j}
F_{ij}^{n_0}(|\vec x_i-\vec y_j|) C_{n_{0_{ij}}}| \le \sum_{i,j}|F_{ij}^{n_0}
(|\vec x_i-\vec y_j|) C_{n_{0_{ij}}}|,
$$
$$
\sum_{i,j}|F_{ij}^{n_0}
(|\vec x_i-\vec y_j|) C_{n_{0_{ij}}}| \le \max_{\vec x_i \in \Delta_1,
\vec y_j \in \Delta_2} |F_{ij}^{n_0}
(|\vec x_i-\vec y_j)||\sum_{i,j}|C_{n_{0_{ij}}}|,
\eqno(A.15)
$$
where the maximum is taken over all possible $\vec x_i \in \Delta_1$
and $\vec y_j \in \Delta_2$,
and
$$\eqalign{
\max |P(AB)-P(A)P(B)| & \sim \max |\sum_{i,j}
F_{ij}^{n_0}(|\vec x_i-\vec y_j|) C_{n_{0_{ij}}}|
\cr &
\le \max_{\vec x_i \in \Delta_1,
\vec y_j \in \Delta_2} |F_{ij}^{n_0}
(|\vec x_i-\vec y_j|)| \max (\sum_{i,j}|C_{n_{0_{ij}}}|), \cr}
\eqno(A.16)
$$
where the maximum on the left hand is taken over all possible
integration ranges. On the right hand of (A.16),
the first maximum is taken over
all possible $\vec x_i \in \Delta_1$
and $\vec y_j \in \Delta_2$, and the second
maximum is in the same sense as the maximum on the left hand.

We know that $|P(AB)-P(A)P(B)|\le 2$
regardless of both the specific values of $F_{ij}$ (within the value range)
and the specific integration ranges for the random variables. Then
the sum $ \sum_{i,j}|C_{n_{0_{ij}}}| $
should converge.
Concerning the approximate behavior when the separation
between $\Delta_1$ and $\Delta_2$ is sufficiently large, we have
$$
\max |P(AB)-P(A)P(B)|\sim C_0\max_{\vec x_i \in \Delta_1,
\vec y_j \in \Delta_2} |F_{ij}^{n_0}
(|\vec x_i-\vec y_j|)|, \eqno(A.17)
$$
where $C_0$ is a finite constant.
By the definition of the mixing rate $\alpha(r_0)$
in (2.9), we have
$$
\alpha(r_0)\sim C_0^{\prime}\max_{|\vec x_i -\vec y_j|\ge r_0}
|F_{ij}^{n_0} (|\vec x_i-\vec y_j|)|, \eqno (A.18)
$$
where $C_0^{\prime}$ is a finite constant and
the maximum is taken over all pairs of ($\vec x_i, \vec y_j$)
with $|\vec x_i -\vec y_j|\ge r_0$.

\par\vfill\end